\documentclass[10pt, a4paper,twocolumn]{article}

\usepackage{booktabs} % For formal tables
\usepackage{qtree} 
\usepackage{xcolor}
\usepackage{caption}
\usepackage{multirow}
\usepackage{subcaption}
\usepackage{graphicx}
\usepackage{amsmath}

\PassOptionsToPackage{hyphens}{url}\usepackage{hyperref}

\makeatother

\begin{document}
\title{Interpreting Social Respect: A Normative Lens for ML Models}

\author{Ben Hutchinson, KJ Pittl, Margaret Mitchell}

\maketitle

\section*{Introduction}

Machine learning is often viewed as an inherently value-neutral process:
statistical tendencies in the training inputs are ``simply''
used to generalize to new examples. However when models impact social
systems such as interactions between humans, these patterns learned by models
have normative implications. It is important that we ask not only ``what
patterns exist in the data?'', but also ``how do we want our system 
to impact people?'' In particular, because minority and marginalized
members of society are often statistically underrepresented in data sets, models
may have undesirable disparate impact on such groups. As such, objectives of
social equity and distributive justice require that we develop tools for both
identifying and interpreting harms introduced by models.

This paper directly addresses the challenge of interpreting how 
human
values are implicitly encoded by deep neural networks, a machine learning
paradigm often seen as inscrutable. Doing so requires understanding how the node
activations of neural networks relate to value-laden human concepts
such as {\sc respectful} and {\sc abusive}, as well as to concepts
about human social identities such as  {\sc gay}, {\sc straight}, 
{\sc male}, {\sc female}, etc. To do this, we present the first application
of Testing with Concept Activation Vectors ({\sc tcav}; \cite{kim2018interpretability})
to models for analyzing human language.

Our contributions are twofold:
1) We present Project Respect, a program and crowdsourcing
    platform for collecting positive  statements about
  marginalized groups.
2) We present experiments into value-driven testing of two
ML models,
using data from Project Respect to create a lens for revealing insights into how layers of
 deep neural networks implicitly encode normative values. 

\begin{figure*}[htbp]
    \centering
\begin{subfigure}{.5\textwidth}
  \centering
    \includegraphics[width=8cm]{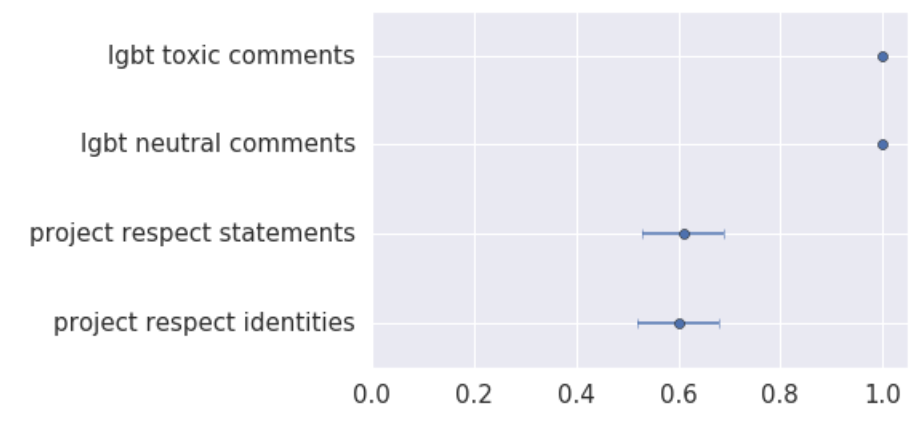}
    \caption{TOXICITY@1}
\end{subfigure}%
\begin{subfigure}{.5\textwidth}
  \centering
    \includegraphics[width=8cm]{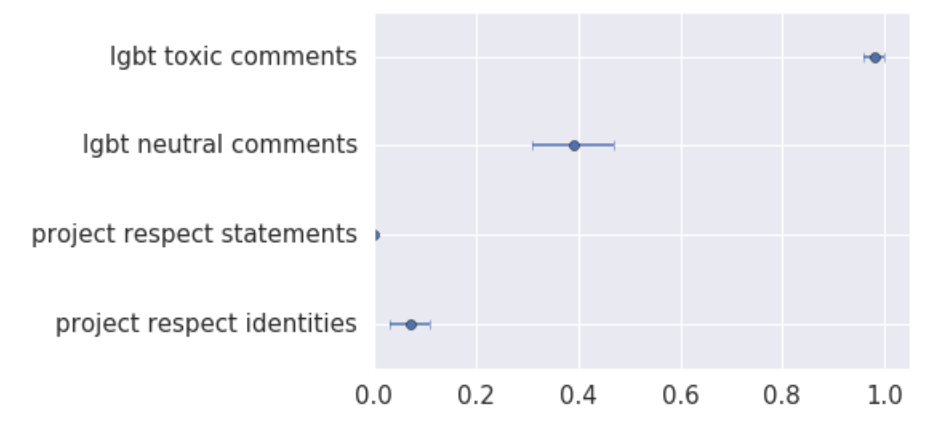}
    \caption{TOXICITY@6}
\end{subfigure}%
\caption{{\sc TCAV} scores for two versions of a model for detecting toxicity in online language, depicting mean {\sc tcav} scores and 90\% confidence intervals.}
\label{fig:results}
\end{figure*}

\section*{Project Respect}

Project Respect
\cite{projectrespect} is a crowdsourcing platform for
collecting positive statements from marginalised communities. The interface enables
community members to enter up to three identity terms (``gay'', ``muslim'', ``transgender'', etc.)
which they would use to describe
themselves. The user then enters positive statements using these identity terms.
The data is collected through the Perspective API \cite{perspectivapi}
and will be open sourced to facilitate further research.

To increase participation,
we conducted outreach and engagement events at several LGBTIQ+ community events, including
Sydney Mardi Gras and San Francisco Pride.
By design,
the data collected by Project Respect has a radically different distribution from language used on the
internet. In particular, whereas much data in online comments
uses words such as ``gay'', ``queer'', and ``transgender''
in abusive or harassing ways \cite{jigsaw}, the data we collect
has language that has positive sentiment.

\section*{Normative model insights}

This works makes a distinction between a {\it descriptive} approach to machine learning, which uses the training data as-is to learn and reproduce likely patterns, and a {\it normative} approach to machine learning, where additional data and constraints are added to define the values we believe ought to hold.

The Perspective API is a tool for improving online conversations by assisting in the
detection of abuse and harassment. It takes a comment
and returns a score between 0 and 1, with 1 indicating a high confidence that the comment
is toxic, i.e. inappropriate. We applied Testing with Concept
Activation Vectors ({\sc TCAV}; \cite{kim2018interpretability})
to versions 1 and 6 of the model, i.e.\ {\sc toxicity@1} and {\sc toxicity@6}, the latter of which had bias mitigation
similar to the techniques described in \cite{perspectivapi}.

{\sc TCAV} models concepts as vectors in spaces defined by internal activations of
the neural network. {\sc TCAV} scores indicates how important the concept is for the model's prediction of toxicity.
For example, the internal representation of a comment can be moved in the direction of the vector representing
a concept, and we can observe whether the models prediction of toxicity increases or decreases. 
%In other words, the {\sc TCAV} score measures the sensitivity of the toxicity prediction to the concept, with 
A score
of 1 for a concept means that the concept is positively associated with toxicity in the model's internal representations.

{\sc tcav} relies on sets of examples
To learn the vector representations of concepts; for our experiments we used 
four sets of examples associated with values.. The first, ``{\sc LGBT} toxic'', contains comments that were randomly sampled from  a large database of online comments that
talked about  
{\sc LGBTIQ+} identities and were identified by human raters as being toxic.
The second, ``LGBT neutral'', was similar but human raters 
identified the comments as non-toxic.
The final two sets of examples embody the normative values in this work: self-identified LGBTIQ+ terms and positive statements about LGBTIQ+ identities collected by Project Respect.
%Inspection
%of the data confirmed that the statements expressed %positive sentiment.  

The two versions of the toxicity model displayed very different
results, as  shown in Figure~\ref{fig:results}.
Whereas version 1 had 
high {\sc  tcav} scores for both toxic  and neutral comments
using {\sc LGBTIQ+} identity terms sampled from the web, version 6
assigned very different {\sc tcav } scores to  
toxic and neutral comments.
The data sets collected using Project Respect had moderately high {\sc tcav} scores
for version, and much lower scores for  version 6. 

%To learn the vector representations of concepts, {\sc tcav} relies on sets of examples. For our experiments we used 
%nine sets of ``non-normative'' examples reflecting actual online language use, as well as four sets of ``normative''
%examples that contained social identity terms.
%The non-normative data sets 
%consisted of comments that were randomly sampled from  a large ($\approx$2M) database of online comments from multiple domains.
%These that 
%contained the social identity terms ``gay'', ``lesbian'', ``homosexual'', ``heterosexual'', ``man'' and ``woman''.
%These samples contained no normative filtering, i.e.\ they contained abusive, positive, and neutral language
%in proportions reflecting the descriptive distributions on the internet.
%As baselines, we included
%comments containing the words ``tree'', ``table'' and ``person''. These 
%The first two ``descriptive'' data sets, ``LGBT negative'' and ``LGBT neutral'',
%consisted two samples from the same dataset of online comments, filtered so that instances both i) contain {
%\sc LGBTIQ+} 
%identity terms, and ii) were judged toxic or non-toxic by human raters, respectively.
%The final two ``normative'' concepts were learned from the identity 
%terms and statements collected by Project Respect.
%Results are  shown in Figure~\ref{fig:results}.

\section*{Discussion and Conclusion}

These results show that the internal representations of the
two neural networks are encoding different kinds of information about the interaction between social
identities and normative values.
Since the techniques for exploring the internal model representations are applied to pre-trained 
models and require just sets of examples, they are easily adaptable to other models and other domains.
They provide a normative lens for understanding the internal representations which lead
models to produce disparate outcomes for different groups.

\bibliographystyle{abbrv}
\bibliography{interpreting_respect}

\end{document}